\documentclass[12pt,english,twoside]{article}

\usepackage[T1]{fontenc}
\usepackage[utf8]{inputenc}
\usepackage[letterpaper, margin=2cm]{geometry}
\usepackage{color}
\usepackage{graphicx}
\usepackage{subfigure}
\graphicspath{{graphics_fold/}}
\usepackage{epstopdf} %Converts EPS to PDF so PDFLaTeX can be used
\usepackage{times}
\usepackage{etoolbox}
\usepackage{amsmath} % assumes amsmath package installed
\usepackage{amssymb}  % assumes amsmath package installed
\usepackage{fancyhdr}
\usepackage{psfrag}
\usepackage{enumitem}
\usepackage{indentfirst}       % indent the first line of the section and subsection
\usepackage{titlesec}          % customizing the section titles
\usepackage{hyperref}
\usepackage{babel}
\usepackage{listings}
\usepackage{booktabs}

\addto\captionsenglish{}

\usepackage[colorinlistoftodos]{todonotes} % Cool To-Do bubbles!
\setlist{nolistsep}

\makeatletter
\renewcommand*{\@biblabel}[1]{\hfill#1.}
\makeatother

\makeatletter
\patchcmd{\headrule}{\hrule}{\color{blue}\hrule}{}{}
\patchcmd{\footrule}{\hrule}{\color{blue}\hrule}{}{}
\makeatother

\fancypagestyle{firstpage}{
\fancyhead{}
\fancyfoot{}
\fancyfoot[LE,RO]{\small \thepage}

}

\makeatletter
\def\maketitle{
  \thispagestyle{firstpage}
\vspace*{-11mm}{\centering\includegraphics[width=0.86\textwidth]{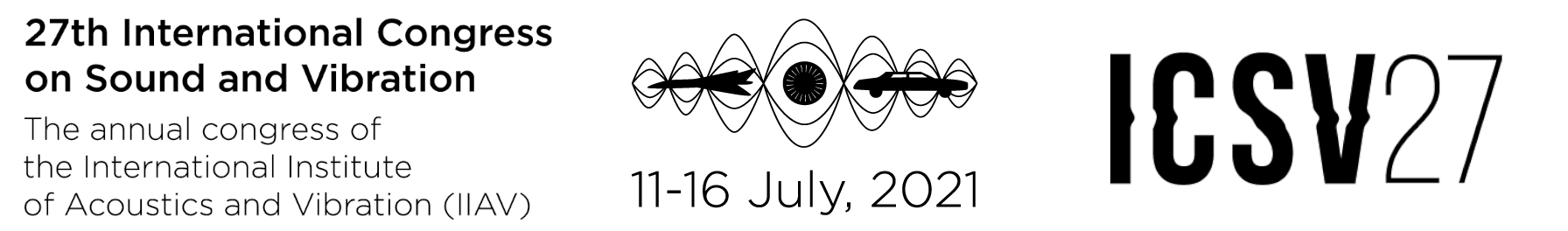}\\}
  {
   \vspace*{-2mm}\fontsize{11}{20}\selectfont\rmfamily{} \noindent {\bf Annual Congress of the International Institute of Acoustics and Vibration (IIAV)}\\

   \vspace*{-7mm}\fontsize{17}{20}\selectfont\sffamily{}  \noindent \MakeUppercase{\textbf{\@title}}

   \vspace*{3mm}\fontsize{14}{20}\selectfont\rmfamily{} \noindent \@author
  }
}
\makeatother

\titleformat{\section}
  {\fontsize{14}{14}\normalfont\sffamily\bfseries}
  {\thesection.}{1em}{}                % insert a point after the section number like the original .doc file

\titleformat{\subsection}
  {\normalfont\sffamily\bfseries}
  {\thesubsection}{1em}{}

\titleformat{\subsubsection}
  {\normalfont\sffamily\small}
  {\textit{\thesubsubsection}}{1em}\textit{{}}

\title{SOUND PRESSURE MINIMIZATION AT THE EAR DRUM FOR IN-EAR ANC HEADPHONES USING A FIXED FEEDFORWARD REMOTE MICROPHONE TECHNIQUE}

\author{Piero Rivera Benois\\
{\small \textit{University of Oldenburg, Oldenburg, Germany\\
e-mail: piero.rivera.benois@uni-oldenburg.de}}\\
Simon Doclo\\
{\small \textit{University of Oldenburg, Oldenburg, Germany\\
email: simon.doclo@uni-oldenburg.de}}}

\author{Piero Rivera Benois$^{1,3}$, Reinhild Roden$^{2}$, Matthias Blau$^{2,3}$ and Simon Doclo$^{1,3}$\\
	{\small \textit{$^{1}$Signal Processing Group, University of Oldenburg, Oldenburg, Germany\\
			$^{2}$Institut für Hörtechnik und Audiologie, Jade Hochschule, Oldenburg, Germany\\
			$^{3}$Cluster of Excellence Hearing4all\\
			e-mail: piero.rivera.benois@uni-oldenburg.de}}}

\begin{document}

\maketitle
\renewcommand{\abstractname}{\vspace{-\baselineskip}} % erase the space between authors and abstract

\begin{abstract}	\noindent
In this paper we consider an in-ear headphone equipped with an external microphone and aim to minimize the sound pressure at the ear drum by means of a fixed feedforward ANC controller. Based on measured acoustic paths to predict the sound pressure generated by external sources and the headphone at the ear drum, the FIR filter coefficients of the ANC controller are optimized for different sound fields. Due to the acoustic feedback path between the loudspeaker and the microphone, a stability constraint based on the Nyquist stability criterion is introduced. Performance degradations due to reinsertions of the headphone and intra-subject variations are addressed by simultaneously optimizing the controller for several measurement repetitions of the acoustic paths. Simulations show that the controller optimized for an ipsilateral excitation produces an attenuation of at least -10 dB that extends approximately to +45° and -65° from the ipsilateral DoA. The controller optimized for a diffuse-field excitation achieves an attenuation of at least -10 dB over a wider range of DoAs on the ipsilateral side, namely +90° to -90°. Optimizing the controller for several measurement repetitions is shown to be effective against performance degradations due to reinsertions and intra-subject variations. 
\end{abstract}

\quad\rule{425pt}{0.4pt}

\section{Introduction}

%ANC in general
Active Noise Control (ANC) applied to headphones aims at minimizing the environmental noise at the listener's ears \cite{simshauser1955}. These devices combine the passive attenuation achieved by the construction material of the headphone and the active attenuation achieved by means of ANC. While passive attenuation is mainly effective in the mid and high frequencies, ANC is mainly effective in the low frequencies. The working principle of ANC is based on the destructive superposition of sound waves. In this context, the sound wave of the environmental noise reaching the listener's ear overlaps with the sound wave generated by the loudspeaker of the headphone. If both sound waves have the same magnitude but opposite phase, the resulting sound pressure is zero. 

%\begin{figure}[!h]
%	\centering
%	\includegraphics[width=3.5cm]{figures/general_overview.png}
%	\caption{In-Ear headphones overview}
%\end{figure}

\begin{figure}[]
	\centering
	\subfigure[][]{
		\includegraphics[width=3.5cm,trim=0 0 0 0,clip]{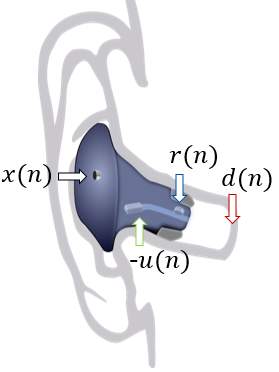}
		\label{fig:general_overview}
	}\hfill
	\subfigure[][]{
		\includegraphics[width=12cm]{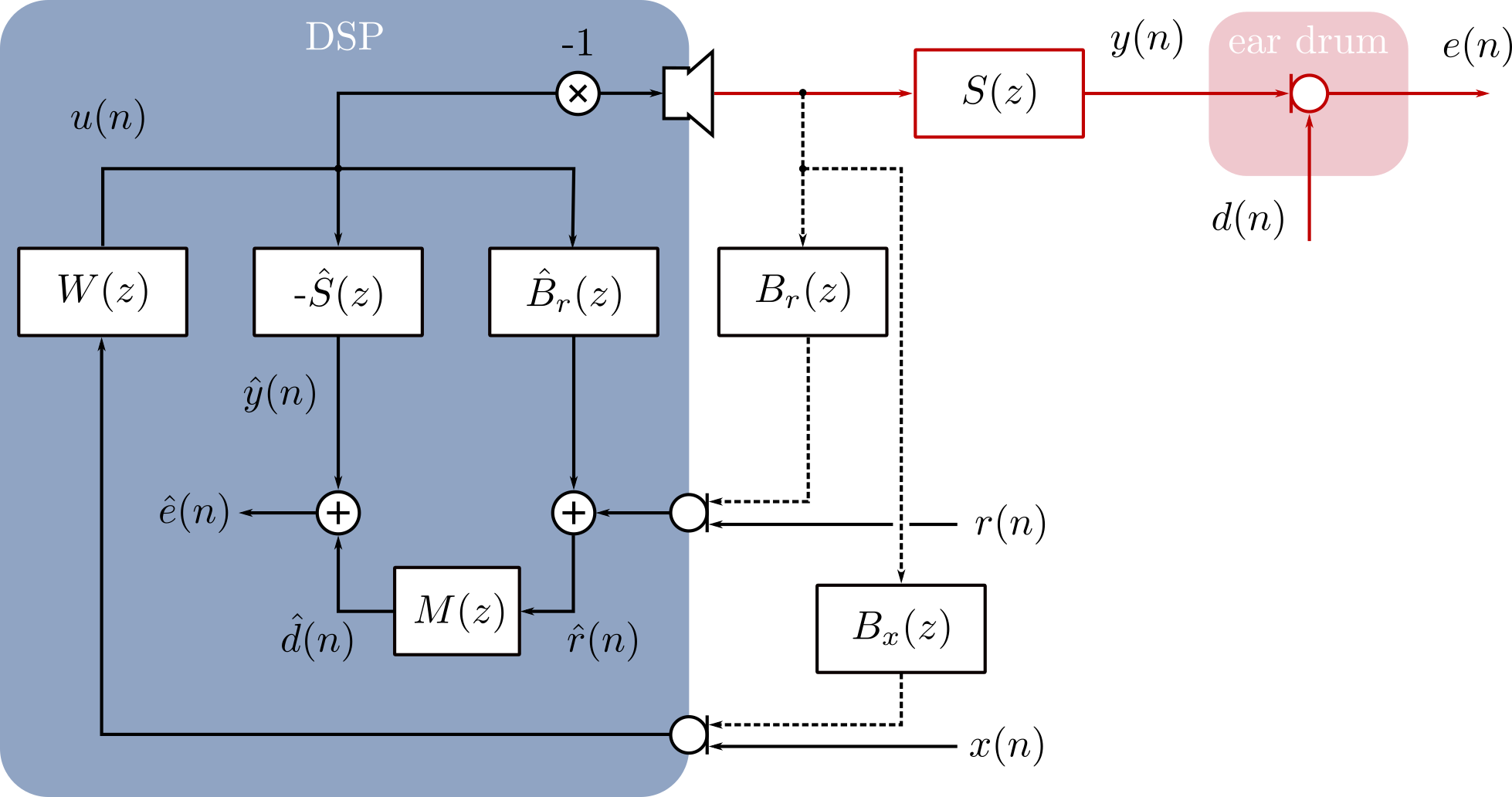}
		\label{fig:fb_vsa}
	}
	\caption{(a) in-ear headphone, where $x(n)$, $r(n)$ and $d(n)$ denote the signals generated by the incident noise arriving at the external microphone, the inner microphone and the ear drum, respectively, and $-u(n)$ denotes the loudspeaker signal generated by the ANC controller. (b) block diagram of an ANC controller implementing a feedforward virtual sensing algorithm using both the external and inner microphones.}
\end{figure}
% overview
%Generally, in-ear ANC headphones are equipped as shown in Fig.\ref{fig:general_overview}. An external microphone measures the incident noise $x(n)$. The noise's sound wave travels further through the device, along the ear canal, and reaches the position of the inner microphone as $r(n)$. Finally, the noise's sound wave reaches the ear drum as $d(n)$. Simultaneous to the sound transmission, the ANC controller (not shown in the picture) generates the control signal $-u(n)$ to drive the loudspeaker and generate a sound wave that destructively overlaps with $d(n)$ at the position of the ear drum. ANC solutions that use the time-advanced signal $x(n)$ for generating $-u(n)$ are called feedforward approaches \cite[Chapter~3]{elliott2001}, while the ones that use $r(n)$ instead are denoted as feedback approaches \cite[Chapter~6]{elliott2001}. Feedforward solutions based on adaptive filter techniques make also use of $r(n)$ for monitoring the optimality of the filter \cite{kuo2000}.  
 
Fig.\ref{fig:general_overview} depicts an in-ear headphone equipped with an external microphone, an inner microphone and a loudspeaker. The wave of the incident noise first arrives at the external microphone as $x(n)$, then travels through the device until it reaches the inner microphone as $r(n)$, and finally reaches the ear drum as $d(n)$. Simultaneously, the ANC controller generates the signal $-u(n)$ to drive the loudspeaker, aiming at generating a sound wave that destructively overlaps with the sound wave of the incident noise. Several ANC approaches have been proposed in the literature \cite{kuosbook}. ANC approaches using the time-advanced external microphone signal to generate the loudspeaker signal are called feedforward approaches, whereas ANC approaches using the inner microphone signal instead are called feedback approaches. When an inner microphone is available, feedforward approaches can exploit the inner microphone signal to monitor the optimality of the ANC filter in minimizing the sound pressure at the inner microphone \cite{elliott2001},

Since the sound pressure at the inner microphone and the ear drum can not be assumed to be the same, virtual sensing algorithms \cite{moreau2008} have been proposed that aim at minimizing the sound pressure at the ear drum. These algorithms make use of transfer functions measured during a calibration stage with a microphone placed at the ear drum, for instance using an audiological probe tube microphone. This microphone is removed for the control stage, but the measured transfer functions are incorporated in the ANC algorithm to make online approximations of the sound pressure at the ear drum. 

Fig.\ref{fig:fb_vsa} depicts the block diagram of the typical feedforward virtual sensing algorithm exploiting the availability of an inner microphone. First, the measured transfer function between the loudspeaker and the inner microphone $\hat{B}_r(z)$ is used to estimate the incident noise at the inner microphone $\hat{r}(n)$. Second, the incident noise at the ear drum $\hat{d}(n)$ is estimated from $\hat{r}(n)$. On the one hand, in the virtual microphone arrangement (VMA) \cite{elliott1992} it is simply assumed that the incident noise at the ear drum is the same as the incident noise at the inner microphone, i.e. $\hat{d}(n)=\hat{r}(n)$ and $M(z)=1$. On the other hand, the remote microphone technique (RMT) \cite{roure1999} assumes that the incident noise at the inner microphone and the incident noise at the ear drum are related by the time-invariant transfer function $M(z)=\Phi_{dr}(z)/\Phi_{rr}(z)$, using which $\hat{d}(n)$ can be estimated from $\hat{r}(n)$. Using the measured transfer function between the loudspeaker and the ear drum $\hat{S}(z)$ the sound pressure at the ear drum $\hat{e}(n)$ can then be estimated by adding the estimated control signal at the ear drum $\hat{y}(n)$ to $\hat{d}(n)$. The estimated sound pressure at the ear drum $\hat{e}(n)$ can then be used to optimize/adapt the feedforward controller $W(z)$ \cite{cheer2018}.

%The virtual microphone arrangement (VMA) \cite{elliott1992} (see Fig.\ref{fig:fb_vsa}) uses the measured transfer function between its loudspeaker and the inner microphone $\hat{B}_r(z)$ to make an approximation of the incident noise at the inner microphone $\hat{r}(n)$. By assuming that the incident noise is approximately the same at the microphone and ear drum $d(n)\approx r(n) \implies M(z)= 1$, an approximation of the incident noise at the ear drum $\hat{d}(n)$ is made. The measured transfer function between its loudspeaker and the ear drum $\hat{S}(z)$ provides an estimation of the control signal at the ear drum. This is subtracted from $\hat{d}(n)$ to approximate the resulting sound pressure at the ear drum $\hat{e}(n)$. This approximated sound pressure is either used for adapting the controller $W(z)$ or to replace $x(n)$ as input for the controller in a feedback approach \cite{pawelczyk2009}. The remote microphone technique (RMT) \cite{roure1999} works in a similar way. However, it abandons the assumption that the incident noise is the same at the inner microphone's and ear drum's positions. Instead it assumes that there is a stable transfer function $M(z)=\Phi_{dr}(z)/\Phi_{rr}(z)$ between $d(n)$ and $r(n)$ that can be used to calculate $\hat{d}(n)$ using $\hat{r}(n)$. Adaptive feedforward controllers use the estimated $\hat{e}(n)$ for adapting $W(z)$ \cite{jung2017}, while the feedback approach in \cite{das2011} uses $\hat{r}(n)$ as input to the controller $W(z)$ and $\hat{e}$ for adapting it.
\begin{figure}[!h]
	\centering
	\includegraphics[width=11cm]{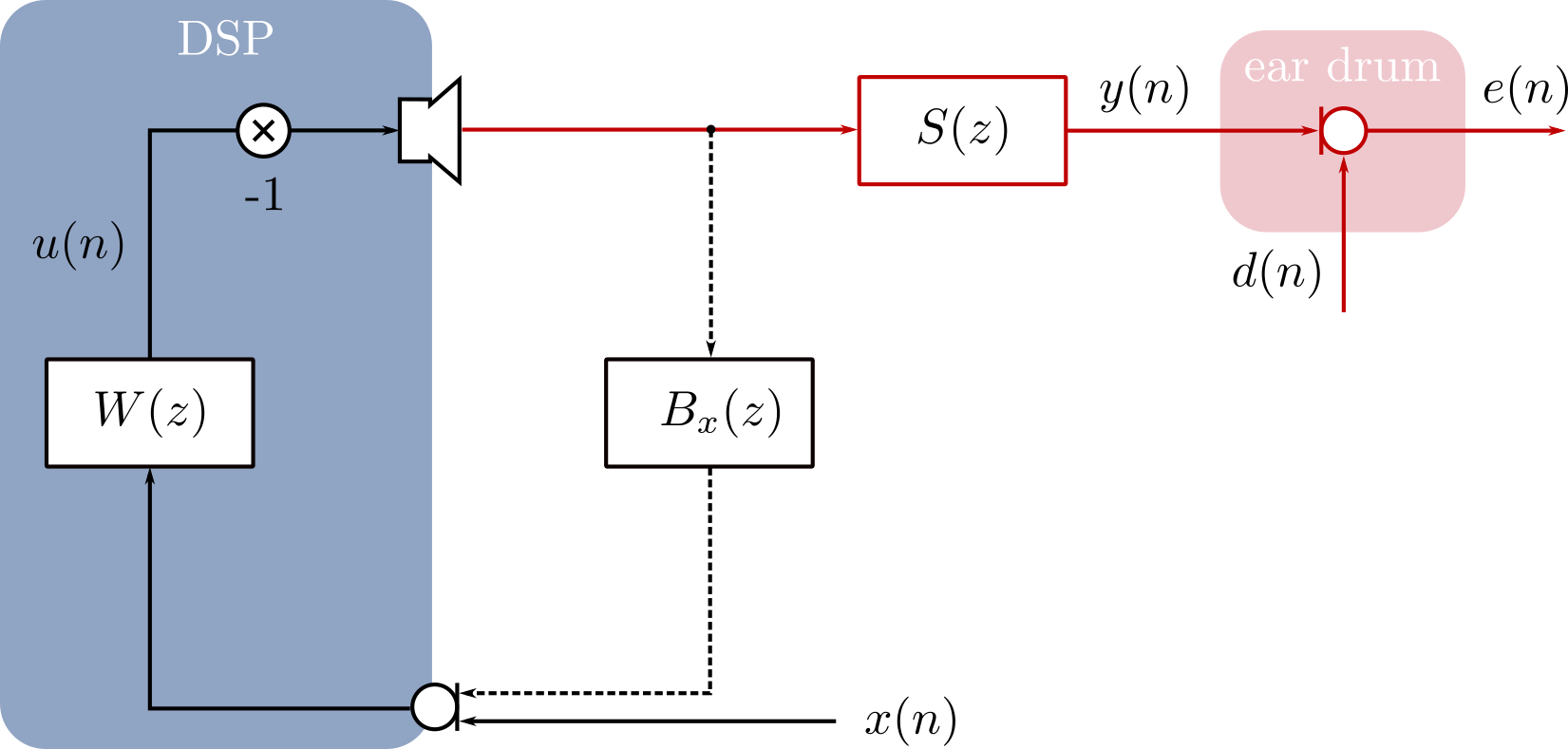}
	\caption{Block diagram of the proposed fixed feedforward remote microphone technique using only the external microphone.}
	\label{fig:fixed_rmt}
\end{figure}

In this paper, we consider an in-ear headphone equipped with only an external microphone and propose a robust design procedure to compute a fixed feedforward filter aiming at minimizing the sound pressure at the ear drum. Similarly as for the RMT, we assume that the incident noise at the external microphone and the incident noise at the ear drum are related by the time-invariant transfer function $\Phi_{dx}(z)/\Phi_{xx}(z)$. Based on measured acoustic transfer functions, the FIR filter coefficients of the ANC controller are optimized subject to a stability constraint, taking the feedback coupling between the loudspeaker and the external microphone into account. We propose to increase the robustness against reinsertion and intra-subject variations by simultaneously optimizing the controller for several measurement repetitions.

%In this paper we are interested in the optimization of a fixed single-input single-output ANC filter to minimize the power spectral density (PSD) at the ear drum. Although its working principle is of feedforward nature, since it makes use of the time-advanced signal $x(n)$, the fact that it only uses one microphone makes it comparable to the feedback VMA in \cite{pawelczyk2009}. However, in our case we consider that the noise at the microphone and at the ear drum are different, but related through the stable transfer function $\Phi_{dx}(z)/\Phi_{xx}(z)$. Hence, the proposed solution makes the same assumptions as the RMT and is, therefore, classified as the fixed-filter variant of it. In the proposed approach, the optimization of the controller $W(z)$ and the estimation of $\Phi_{dx}(z)/\Phi_{xx}(z)$ are implicitly combined and performed during the calibration stage. The feedback coupling between its loudspeaker and microphone is addressed similarly to \cite{pawelczyk2009} by a Nyquist stability criterion analysis. However the proposed stability constraint ensures not only a gain margin but also a phase margin to the system. Hence, a more robust stability can be ensured. Intra-subject and reinsertion variability are addressed by optimizing the controller for several measurement repetitions. 

In Section\,\ref{sec:fix_ff_rmt} the proposed system is presented together with the cost function and stability constraint used for optimizing the controller $W(z)$. In Section\,\ref{sec:results} simulations results based on measured transfer functions are used to validate the proposed approach.

\section{Fixed Feedforward Remote Microphone Technique}
\label{sec:fix_ff_rmt}
 %The most important difference to Fig.\,\ref{fig:fb_vsa} lies on the fact, that for a fixed feedforward controller the monitoring signal $r(n)$ is not used. Therefore, the calculation of the fixed controller and the estimation of $M(z)$ has to be done jointly during the calibration stage.
 
The block diagram of the proposed system is presented in Fig.\,\ref{fig:fixed_rmt}. The incident noise $x(n)$ is measured by the external microphone. This time-advanced signal is filtered by the fixed controller $W(z)$ to calculate the control signal $-u(n)$, which is used to drive the loudspeaker. The generated sound wave travels on the one hand through the feedback path $B_x(z)$ to the external microphone and on the other hand through the secondary path $S(z)$ to the ear drum. The signal $e(n)$ at the ear drum is the sum of the control signal $y(n)$ and the incident noise $d(n)$ at the ear drum. The objective is to design the controller $W(z)$ such that the PSD of the signal $e(n)$ is minimized. The PSD can be written as
\begin{equation}
\label{eq:Phi_ee_ff}
\Phi_{ee}(f)=\bigg(1- \frac{|\Phi_{dx}(f)|^2}{\Phi_{dd}(f)\Phi_{xx}(f)}\bigg)\Phi_{dd}(f)+\bigg|\frac{\Phi_{dx}(f)}{\Phi_{xx}(f)}-\frac{W(f)}{1+W(f)B_x(f)}S(f)\bigg|^2\Phi_{xx}(f),
\end{equation}
where $\Phi_{dx}(f)$ is the cross-power spectral density (CPSD) of $d(n)$ and $x(n)$, and $\Phi_{dd}(f)$ and $\Phi_{xx}(f)$ are the PSDs of the respective signals. The left-hand addend determines the minimum achievable $\Phi_{ee}(f)$, which is equal to zero when the magnitude squared coherence between $d(n)$ and $x(n)$ is equal to 1. The right-hand addend defines the scope of optimization of the controller $W(z)$. Here, the controller and the feedback path $B_x(z)$ build a transfer function, which in series connection with the secondary path $S(z)$ should approximate the transfer function $\Phi_{dx}(z)/\Phi_{xx}(z)$ as well as possible.

%In this context we consider that the transfer function  $\Phi_{dx}(z)/\Phi_{xx}(z)$ is mainly determined by the headphones and subject, and less by the incident noise field. Therefore, the approach that we take to optimize $W(z)$ is to prioritize the reinsertion and intra-subject variability over the ones that different sound fields would produce. We do this by optimizing the controller for several measurement repetitions considering a generic or approximated sound field. Following this approach, the finite impulse response of the controller $\textbf{w}$ is optimized using cost function formulated in the DFT domain as

\begin{figure}[]
	\centering
	\subfigure[][]{
		\includegraphics[width=6.25cm,trim=0 0 0 0,clip]{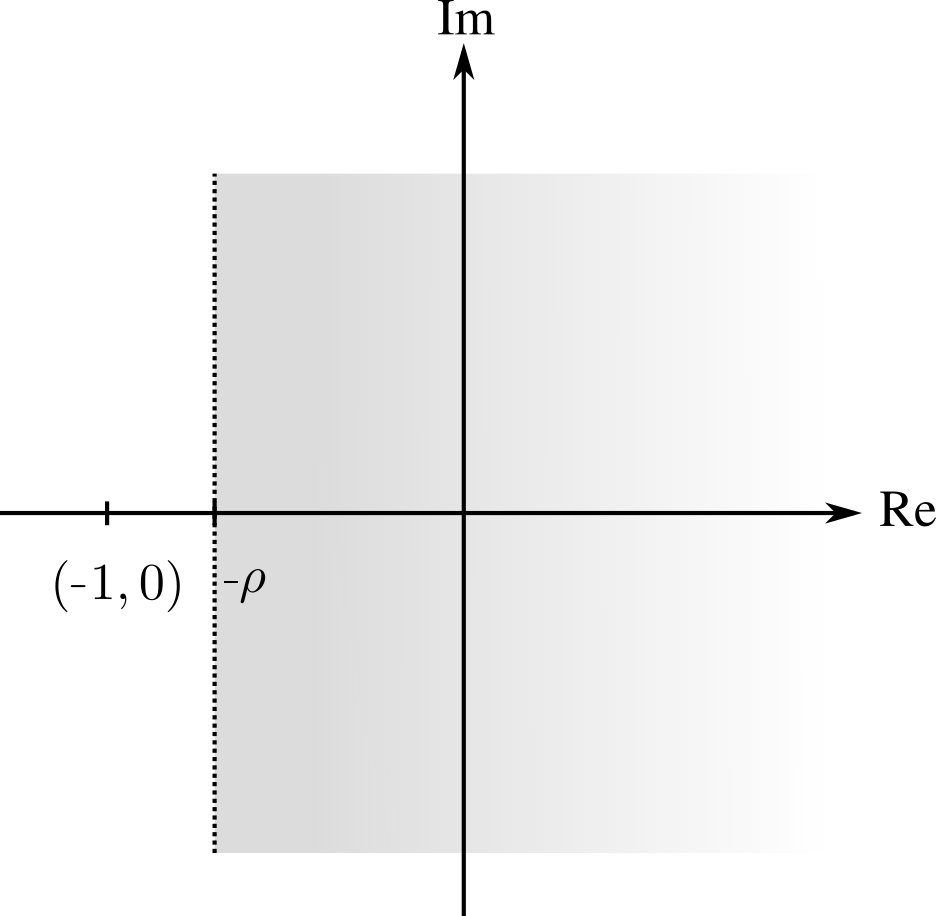}
		\label{fig:vertical_boundary}
	}\hfill
	\subfigure[][]{
	\includegraphics[width=6.25cm]{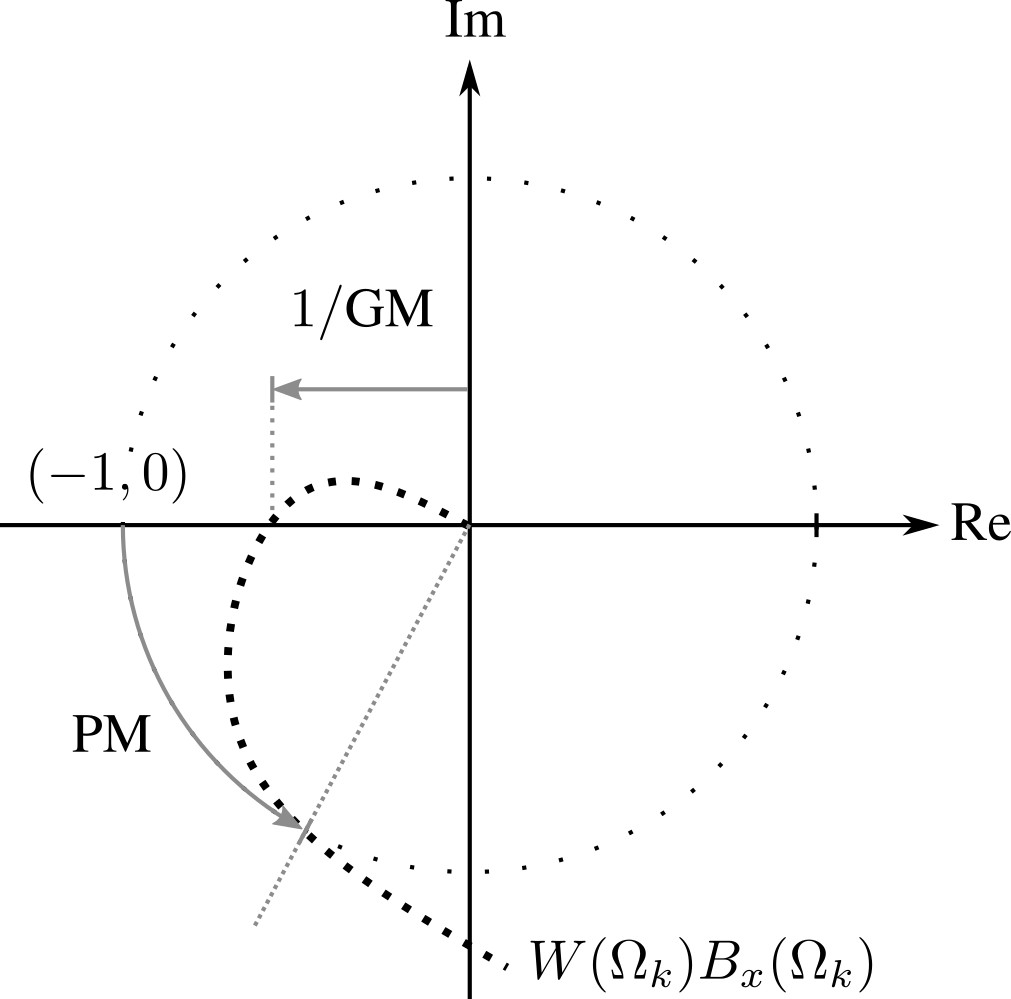}
		\label{fig:gm_and_pm}
	}
	\caption{In (a) the solution space (in gray) generated by a vertical boundary as stability constraint, and in (b) an example of the gain and phase margins from the contour $W(\Omega_k)B_x(\Omega_k)$.}
	\label{fig:nominal_stability}
\end{figure}

Since the signal at the ear drum can not be directly measured during operation, in this work we will use acoustic transfer functions that were measured during a calibration stage to determine the transfer function $\Phi_{dx}(z)/\Phi_{xx}(z)$ and the secondary path $S(z)$. It should be realized that the transfer function $\Phi_{dx}(z)/\Phi_{xx}(z)$ depends on the position of the headphone, the subject and the incident noise field. Aiming at increasing the robustness of the controller against reinsertions of the headphone and intra-subject variations, we propose to simultaneously optimize the controller for several measurement repetitions. Assuming an FIR filter for the controller $W(z)$, the filter coefficients $\textbf{w}$ are calculated by minimizing the regularized cost function in the DFT domain
\begin{equation}
\label{eq:cost_function}
\min_\textbf{w} \sum_{r=0}^{R-1}\sum_{k=0}^{L_\text{DFT}/2}\bigg|\frac{\hat{\Phi}_{dx}(\Omega_k,r)}{\hat{\Phi}_{xx}(\Omega_k,r)}-\frac{W(\Omega_k)}{1+W(\Omega_k)B_x(\Omega_k,r)}\hat{S}(\Omega_k,r)\bigg|^2\hat{\Phi}_{xx}(\Omega_k,r)+\beta \big|W(\Omega_k)\big|^2,
\end{equation}
where $W(\Omega_k)$ denotes the frequency response of the filter $W(z)$ at the normalized frequency $\Omega_k=2\pi k/L_\text{DFT}$, $L_\text{DFT}$ the DFT length, $R$ the number of measurement repetitions and $\beta$ the Tikhonov regularization parameter. The CPSD $\hat{\Phi}_{dx}(\Omega_k,r)$ and the PSD $\hat{\Phi}_{xx}(\Omega_k,r)$ are calculated from the measured signals used during the $r^\text{th}$ measurement of the calibration stage. Similarly, $\hat{S}(\Omega_k,r)$ denotes the secondary path between the loudspeaker and the ear drum for the $r^\text{th}$ measurement during the calibration stage. Depending on the headphone design and construction materials, the feedback path can sometimes be neglected, i.e. $B_x(\Omega_k,r) \approx 0$. However, in the general case either the system requires a feedback control algorithm that works in parallel with the ANC algorithm \cite{akhtar2007}, or the optimization needs to include a stability constraint that ensures that the resulting controller yields a stable system. Inspired by \cite{pawelczyk2009}, we propose to use a frequency-dependent constraint based on the Nyquist stability criterion 
\begin{equation}
\label{ineq:nominal_stability}
\big|W(\Omega_k,r_0)B_x(\Omega_k,r_0)\big|^2 - \big|W(\Omega_k)B_x(\Omega_k,r_0)+2\rho\big|^2 < 0 \text{,   } \forall \Omega_k,
\end{equation}
where $\rho$ denotes the robustness parameter and $r_0$ is chosen as the measurement for which the secondary path generates the least restrictive multiplicative uncertainty over frequencies between 100\,Hz and 12\,kHz \cite{skogestad2005}. The constraint in (\ref{ineq:nominal_stability}) ensures that the solution space in the complex plane of the Bode plot is restricted to the right-hand side of a vertical boundary determined by $\rho$, as shown in Fig.\,\ref{fig:vertical_boundary}. This avoids that the contour drawn by the open-loop transfer function frequency response $W(\Omega_k)B_x(\Omega_k,r_0)$ encircles the Nyquist point at $(-1,0)$, see Fig.\,\ref{fig:gm_and_pm}. As a direct result, the gain and phase margins satisfy
\begin{equation}
\frac{1}{\text{GM}} \leq \rho 
\end{equation}
and
\begin{equation}
\text{PM} \geq \arccos(\rho),
\end{equation}
respectively. Both margins protect the stability of the system against discrepancies in magnitude and phase between the measured feedback path $B_x(\Omega_k,r_0)$ and the real feedback path $B_x(\Omega_k)$. For instance, by setting $\rho=0.8$, gain and phase margins of $\text{GM} = 1.25$ and $\text{PM}\approx37^\circ$ can be produced. The proposed constrained nonlinear optimization problem can then be solved using the interior-point or sequential quadratic programming algorithms implemented, for example, in the MATLAB function \texttt{fmincon()}. 

\section{Results}
\label{sec:results}

In this section the measured acoustic paths relative to $R=7$ measurement repetitions and one subject from the database in \cite{denk2021} are used. Three types of sound fields are considered as possible calibration noise fields: (a) a cylindrical-isotropic field generated by uncorrelated broadband noise sources, which are uniformly distributed around the subject, such that the direction of arrival (DoA) angular resolution is equal to $7.5^\circ$; (b) a single noise source standing in free field at the ipsilateral (right-hand, DoA = $270^\circ$) side; and (c) a single noise source standing in free field at the contralateral (left-hand, DoA = $90^\circ$) side. For each measurement repetition these calibration noise fields are generated with a sampling frequency of $f_s=44.1$\,kHz for a duration of $4$ seconds. From the simulated external microphone and ear drum signals biased auto- and cross-correlations $\hat{\phi}_{xx}(\tau,r)$ and $\hat{\phi}_{dx}(\tau,r)$ are estimated for a lag span of $\pm\tau_\text{max}=L_s+L_w-1$, where $L_s$ is the impulse response length of the secondary path $\hat{S}(z)$ and $L_w$ is the impulse response length of the ANC filter $W(z)$. In our experiments we have used $L_w=512$ and $L_s=712$. The biased auto- and cross-correlations are DFT-transformed using a DFT length of $L_\text{DFT}=8192$. The Tikhonov regularization parameter in (\ref{eq:cost_function}) is set to $\beta = 0.01 \cdot \hat{\phi}_{xx}(\tau=0,r_0)$ and the robustness parameter in (\ref{ineq:nominal_stability}) is set to $\rho=0.8$. 

%estimated auto- and cross-correlations $\hat{\Phi}_{dx}(\Omega_k,r)$ and $\hat{\Phi}_{xx}(\Omega_k,r)$

In a first experiment, the influence of the calibration noise field on the attenuation performance in free field is investigated. Here, one reinsertion is considered ($R=1$) and one controller is optimized for each calibration noise field, i.e. $\boldsymbol{w}^\text{diff}$, $\boldsymbol{w}^\text{ipsi}$ and $\boldsymbol{w}^\text{contra}$. The system is excited in free field sequentially, during 4\,s for each DoA, using an angular resolution of $7.5^\circ$. This is done either with ANC on or off. When ANC is on, the band-limited signal power of the remaining sound pressure at the ear drum $e(n)$ can be estimated as
\begin{equation}
P_\text{on}(f_\text{low}, f_\text{high}) = \frac{1}{k_\text{low}- k_\text{high}+1}\sum_{k = k_\text{low}}^{k_\text{high}} \Phi_{ee}(\Omega_k),
\end{equation}
where $k_\text{low}$ is the highest index closest to the lowest frequency of interest $f_\text{low}=100$\,Hz, and $k_\text{low}$ is the lowest index closes to the highest frequency of interest $f_\text{low}=4$\,kHz. Similarly, when ANC is off, the band-limited power of the sound pressure at the ear drum can be estimated as
\begin{equation}
P_\text{off}(f_\text{low}, f_\text{high}) = \frac{1}{k_\text{low}- k_\text{high}+1}\sum_{k = k_\text{low}}^{k_\text{high}} \Phi_{dd}(\Omega_k),
\end{equation}
The ANC performance, i.e. the band-limited attenuation, is calculated in dB by
\begin{equation}
A(f_\text{low}, f_\text{high}) = 10\cdot \log_{10}\left(\frac{P_\text{on}(f_\text{low}, f_\text{high})}{P_\text{off}(f_\text{low}, f_\text{high})}\right)
\end{equation}
for all evaluated DoAs in the horizontal plane. 

\begin{figure}[]
	\centering
	\subfigure[][]{
		\includegraphics[width=9cm,trim=0 0 0 0,clip]{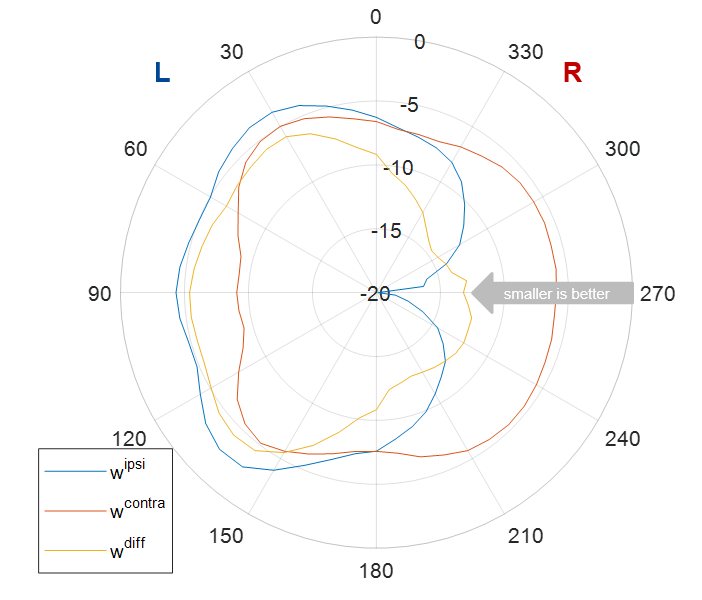}
		\label{fig:att_doa}
	}\hfill
	\subfigure[][]{
		\includegraphics[width=8cm]{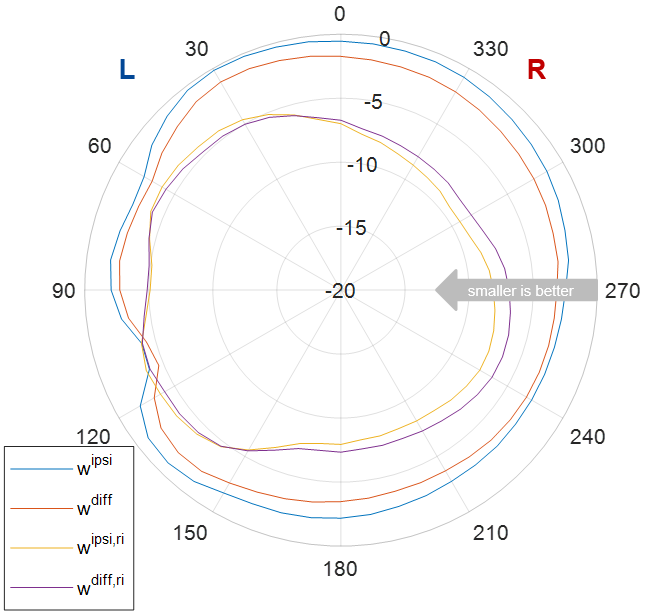}
		\label{fig:att_doa_reinsert}
	}
	\caption{Band-limited attenuation in dB as a function of the direction of arrival (a) for the insertion used during calibration and (b) for a different inserion than used during calibration. The attenuation is calculated for the frequency range of 100\,Hz-4\,kHz.}
\end{figure}

Fig.\,\ref{fig:att_doa} shows the attenuation as a function of the direction of arrival for the three considered controllers $\boldsymbol{w}^\text{diff}$, $\boldsymbol{w}^\text{ipsi}$,  $\boldsymbol{w}^\text{contra}$ for the same headphone insertion that was used to compute the controllers. The controller $\boldsymbol{w}^\text{ipsi}$ optimized for the ipsilateral side (DoA$=270^\circ$) shows a large DoA dependency. It achieves -20\,dB of attenuation when the DoA matches the DoA of the calibration noise field and also at least -10\,dB for the DoAs that extend approximately to $+45^\circ$ and $-65^\circ$ from the ipsilateral DoA. For the controller $\boldsymbol{w}^\text{contra}$ optimized for the contralateral side (DoA$=90^\circ$) similar attenuations as for $\boldsymbol{w}^\text{ipsi}$ can not be achieved for all directions, probably because for the DoAs in the contralateral half hemisphere the performance is limited by a low causality margin, as documented in similar studies \cite{serizel2013}, and not by deviations in the expected DoA. The controller $\boldsymbol{w}^\text{diff}$ optimized for a diffuse field shows an attenuation of at least -10\,dB for the whole ipsilateral half hemisphere. This can probably be explained by the shadowing effect of the human head, which weights the DoAs of the ipsilateral side more.

In the second experiment, Fig.\,\ref{fig:att_doa_reinsert} shows the attenuation achieved with the controllers $\boldsymbol{w}^\text{ipsi}$ and $\boldsymbol{w}^\text{diff}$ from the previous experiment for a different headphone reinsertion than the one used to compute the controllers. It can be seen that after reinsertion an attenuation of at most -4\,dB can be achieved. Moreover, even when using $\boldsymbol{w}^\text{ipsi}$ and exciting the system from the ipsilateral DoA, i.e. when the calibration noise field used in the optimization matches the real sound field, still not more than -2.5\,dB of attenuation can be achieved. A similar attenuation is produced by $\boldsymbol{w}^\text{diff}$. By comparing these two curves with the corresponding ones in Fig.\,\ref{fig:att_doa}, it can be clearly observed that reinsertion variability plays a more dominant role than the mismatch between the calibration noise field and the real noise field. However, if the controllers are optimized using $R=7$ insertions and the evaluation is again performed with the unseen case, both controllers $\boldsymbol{w}^\text{ipsi,ri}$ and $\boldsymbol{w}^\text{diff,ri}$ significantly improve the attenuation in the ipsilateral half hemisphere, with $\boldsymbol{w}^\text{ipsi,ri}$ achieving slightly better results than $\boldsymbol{w}^\text{diff,ri}$. Moreover, the optimization using multiple insertions achieves an improvement of up to -7\,dB of attenuation.

\begin{figure}[!h]
	\centering
	\includegraphics[width=17.5cm]{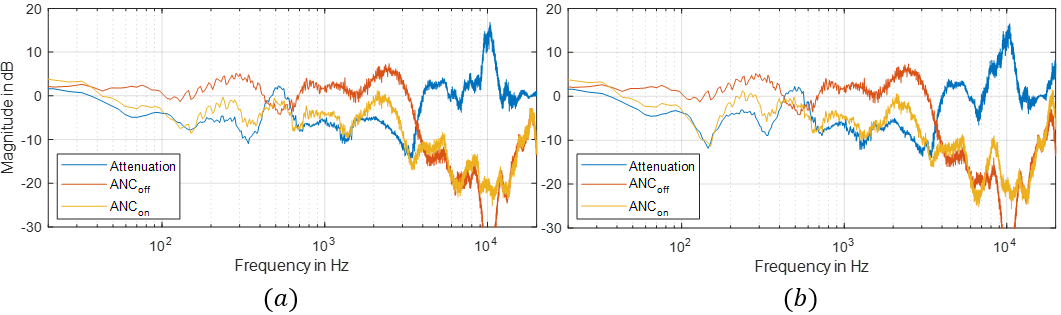}
	\caption{Frequency-dependent PSD at the ear drum with ANC off and on, and frequency-dependent attenuation for (a) robust controller $\boldsymbol{w}^\text{diff,ri}$ optimized for an ipsilateral free-field excitation, and (b) robust controller $\boldsymbol{w}^\text{ipsi,ri}$ optimized for a diffuse-field excitation.}
	\label{fig:att_over_freq}
\end{figure}

In the third experiment, for the robust controllers $\boldsymbol{w}^\text{diff,ri}$ and $\boldsymbol{w}^\text{ipsi,ri}$ from the previous experiment Fig.\,\ref{fig:att_over_freq} shows the frequency-dependent PSD at the ear drum with ANC off and on and the resulting attenuation, for a diffuse-field excitation for a different headphone insertion that the one used to compute the controllers. It can be corroborated that, similarly as in Fig.\,\ref{fig:att_doa_reinsert}, the performance of both controllers is similar. A relatively broad attenuation bandwidth from 40\,Hz to 4\,kHz is achieved, although, around 530\,Hz the attenuation performance drops and an amplification of approx. 2\,dB occurs. At this frequency a drop in the coherence between $d(n)$ and $x(n)$ was observed and an effect of the reinsertion could be dismissed. For frequencies higher than 4\,kHz, it can be seen that amplifications occur. However, this amplifications occur in a frequency region where the headphone provides a large passive attenuation. Due to the similarity between the performance of both considered controllers $\boldsymbol{w}^\text{diff,ri}$ and $\boldsymbol{w}^\text{ipsi,ri}$, it can be argued that during the calibration process the reinsertions could be performed with either calibration noise fields. The ipsilateral single-source free field required by $\boldsymbol{w}^\text{ipsi,ri}$ could be conveniently approximated by a single loudspeaker placed at the ipsilateral side of the headphone. 

\section{Conclusions}
\label{sec:conclusions}
In this paper we investigated a robust design procedure based on the remote microphone technique to calculate a fixed feedforward ANC filter for in-ear headphones in which only the external microphone is available. Assuming an FIR filter for the controller, we optimize the filter coefficients to minimize the sound pressure at the ear drum by making use of transfer functions measured during a calibration stage by means of a microphone placed at the ear drum and a calibration noise field. Aiming at increasing the robustness of the controller against reinsertions of the headphone, we propose to simultaneously optimize the filter coefficients for several measurement repetitions of the acoustic transfer functions. In order to account for the feedback coupling between the loudspeaker and the external microphone, a stability constraint is included in the optimization to ensure that the controller yields a stable system.

Simulations using measured acoustic transfer functions have shown that reinsertions have a much larger effect on the attenuation than the calibration noise field used to optimize the controllers. Simultaneously optimizing for several measurement repetitions has shown to be an effective countermeasure against this type of variability. Using the proposed robust design procedure similar results could be achieved by using either a diffuse-field or a single-source free-field excitation during the calibration. Hence, the ideal setup for the calibration stage would only require a single loudspeaker placed at the ipsilateral side, with which the measurement repetitions could be sequentially performed.
% Based on these similarities it could be concluded that the method could be applied using a cheaper setup if during the calibration stage a single loudspeaker placed at the ipsilateral side is used could achieve comparable results in the future.
\section{Acknowledgments}
This research was funded by the Deutsche Forschungsgemeinschaft (DFG, German Research Foundation) – Project-ID 352015383 – SFB 1330 C1.

\bibliographystyle{icsv_bib}
\bibliography{references}

\end{document}